\documentclass[twocolumn,prl]{revtex4}
\usepackage{epsfig}

\usepackage{graphicx}
\usepackage{amsmath}
\usepackage{epstopdf}

\begin{document}

\title{Evidence of growing spatial correlations during the aging of glassy glycerol}
\author{C. Brun$^1$}
\author{F. Ladieu$^{1 \star}$}
\author{D. L'H\^ote$^{1}$}
\author{G. Biroli$^2$}
\author{J-P. Bouchaud$^3$}
\email{francois.ladieu@cea.fr}

\affiliation{$^1$ SPEC/SPHYNX (CNRS URA 2464), DSM/IRAMIS CEA Saclay, Bat.772, F-91191 Gif-sur-Yvette  France}
\affiliation{$^2$ Institut de Physique Th\'eorique, CEA, (CNRS URA 2306), 91191 Gif-sur-Yvette, France}
\affiliation{$^3$ Capital Fund Management, 6, Bd. Haussmann, 75009 Paris, France}

\date{\today}

\begin{abstract}
We have measured, as a function of the age $t_a$, the aging of the nonlinear dielectric susceptibility $\chi_3$ of glycerol below 
the glass transition. Whereas the linear susceptibility can be accurately accounted for in terms of an age dependent relaxation 
time $\tau_{\alpha}(t_a)$, this scaling breaks down for $\chi_3$, suggesting an increase of the amplitude of $\chi_3$. This is a strong indication that the number $N_{corr}$ of 
molecules involved in 
relaxation events increases with $t_a$. For $T=0.96 \times T_g$, we find that $N_{corr}$ increases by $\sim 10 \%$ when $t_a$ varies from $1\mathrm{ks}$ to $100\mathrm{ks}$. 
This sheds new light on the relation between length scales and time scales in glasses.
\end{abstract}

\maketitle

An old plastic ruler under tension has a longer length than a newly made one \cite{Struick}. This is 
a striking illustration of the \textit{aging} phenomenon, the hallmark of the physics of glasses. 
The physical properties of an aging system depend on the time $t_a$ elapsed since the material has fallen out of equilibrium; 
i.e., since the glass transition has been crossed. Understanding aging is of paramount importance 
\cite{Struick,Ber11}, both from a fundamental and a practical point of view (many daily-life materials do in fact age).  
Yet, there is no universally accepted theoretical description of the basic mechanisms of aging, 
although many scenarii have been proposed (see \cite{NMT,Lub04,Vin07,Leiden,Ber11}).

One of the most distinctive features of aging is the increase of the physical relaxation time $\tau_{\alpha}$ with the age $t_a$ \cite{Struick,Vin07,Leh98,Yar03,Lun05,Shi05,Hec10,Ric10}. In the case of spin-glasses, this increase has been rather
convincingly attributed to the growth of the number $N_{corr}$ of cooperatively relaxing spins. Both  simulations \cite{Yoshino} and experiments \cite{Bert04} are compatible with this scenario, and allow one to estimate the 
dynamical growth law $N_{corr}(t_a)$. The situation is much less clear for most other glassy systems, either experimentally or numerically \cite{Yoshino,Weeks12}. In fact, while there are only two simulations \cite{Parisi,Castillo} reporting the growth of a dynamical correlation length during the aging of model glasses, and few analytical studies \cite{nandi,Bou05,Yoshino}, 
there is to our knowledge {\it no available experimental result} for real glass-formers. The last decade has witnessed an outburst of activity on dynamical heterogeneities and on the determination of the size $N_{corr}$ of dynamically correlated molecules in glasses \cite{Leiden,Ber11}, but almost all these studies have been confined to {\it equilibrated} systems. Whereas a compelling positive correlation between $N_{corr,eq}(T)$ and the equilibrium relaxation time $\tau_{\alpha}(T)$ has been established above the glass temperature $T_g$ (see \cite{Leiden} and refs. therein), its aging counterpart has not been investigated experimentally. The aim of the present study is to extend to the aging regime the experimental determination of $N_{corr}$ that relies on the cubic nonlinear dielectric susceptibility \cite{Cra10,Bru11}. We will report, for the first time, clear experimental evidence of the growth of the size of dynamically correlated regions during the aging of glycerol -- a prototypical glass former. 

As argued in \cite{Bou05},  non-linear susceptibilities are the ideal gambits that elicit the growth of amorphous order in glassy systems. 
Whereas linear susceptibilities (dielectric, magnetic, elastic, etc.) are blind to amorphous order and dynamical correlations, the equilibrium 
cubic nonlinear dielectric susceptibility $\chi_3$ of deeply supercooled glass formers is given, at temperature $T$, by \cite{Bou05}:
\begin{eqnarray}\label{eq1}
\chi_3(\omega,T)& \approx& Z(T) N_{corr,eq}(T) {\cal H}(\omega \tau_{\alpha}(T))\\
Z(T) & \equiv & \frac{\epsilon_0 \Delta \chi_1^2(T) a^3(T)}{k_BT},
\end{eqnarray}
where $\omega=2\pi f$ is the angular frequency, $k_B$ the Boltzmann constant, $\epsilon_0$ the vacuum permittivity, $a^3(T)$ the molecular volume, and $\Delta \chi_{1}(T)= \chi_1(\omega = 0,T) - \chi_1 (\omega 
\gg \tau_{\alpha}^{-1})$ is the contribution to the static linear susceptibility of the degrees of freedom associated with the
glass transition. In Eq. (\ref{eq1}), ${\cal H}(u)$, with $u=\omega \tau_\alpha(T) \equiv f/f_{\alpha}(T)$, is a complex scaling function which goes to zero both for small and large arguments, and peaks in-between. Eq. (\ref{eq1}) can be fully justified within the Mode-Coupling Theory of glasses \cite{Tar10}; it has been confirmed experimentally in \cite{Cra10,Bru11}, and used to extract precise estimates of $N_{corr,eq}(T)$ in equilibrium. In the aging regime, 
it is natural to conjecture \cite{Bou05} that the above expression remains valid with $\tau_\alpha(T) \longrightarrow \tau_\alpha(t_a)$ and $N_{corr,eq}(T) \longrightarrow N_{corr}(t_a)$, therefore allowing one to infer information about the growing of $N_{corr}$ during aging. Strictly speaking, such a simple substitution is too naive: one expects on general ground that (a) the scaling function 
${\cal H}$ should also be replaced by a different scaling function $\tilde{\cal H}$; and (b) the prefactor $Z(T)$ might itself acquire an 
age dependence: the value of both $\Delta \chi_1$ and $a$ could evolve with age, and the temperature $T$ should in principle \cite{Bou05} 
be replaced by an 
``effective'' temperature $T_{eff}(t_a)$ that encodes the possible deviations to the equilibrium fluctuation-dissipation theorem \cite{Cug97,Gri99,Sch11}. However, our experiments are ``weakly'' out of equilibrium, since they reach equilibrium eventually. In this case we expect that the scaling assumption Eq. (\ref{eq1}) generalized to the aging regime, with $\tilde{\cal H}={\cal H}$ and $Z(t_a)=Z(T)$,  holds to a very good approximation. 
Our strategy will therefore be the following: \textit{(i)} since the linear susceptibility does {\it not} depend on $N_{corr}$, its age dependence should only come from that of $\tau_{\alpha}(t_a)$. Indeed, we will establish that $\chi_{1}''(\omega, t_a) \approx {\cal G}''(\omega \tau_{\alpha}(t_a))$, where ${\cal G}''$ is the equilibrium scaling function (see Fig. \ref{fig3}). This allows us to determine $\tau_{\alpha}(t_a)$ directly; \textit{(ii)} by waiting long enough (i.e. $t_a = 200 \mathrm{ks}$) we measure the equilibrium non-linear susceptibility $\chi_3(\omega,T)$ for various frequencies (see Fig. \ref{fig3}), thereby allowing one to obtain the scaling function ${\cal H}(u)$;  \textit{(iii)} inserting these informations into Eq. (\ref{eq1}) in the
 aging regime, we can deduce the age dependence of $N_{corr}$ (up to the assumption that  $Z(t_a)=Z(T)$, see Fig. \ref{fig4}) \cite{Tri}.
 
{\it Experiments.} Ultrapure glycerol was purchased from VWR and placed in our dielectric setup described in Refs \cite{Cra10,Bru11,Thi08}. 
Glycerol was the dielectric layer of a capacitor made with stainless 
steel electrodes separated by a $8.25\mathrm{\mu m}$ thick Mylar$\copyright$ ring. All the 
aging quantities were measured with the same $T$ quench: The sample was first set to $196 \mathrm{K} \approx T_g+8\mathrm{K}$ (where $\tau_{\alpha} \approx 1\mathrm{s}$) during $1 \mathrm{hr}$, then it was cooled, \textit{without any undershoot}, to the working temperature $T=180.1\mathrm{K}$ (or $T=182.7\mathrm{K}$) in $1.8\mathrm{ks}$, and 
finally $T$ was kept constant within a $\pm 70\mathrm{mK}$ interval during $200\mathrm{ks}$. We used a high harmonic purity a.c. field of amplitude $\le 3\mathrm{MVrms/m}$ to measure, separately, $\chi_{3}(\omega,t_a)$ as well as $\chi_1(\omega,t_a)$ and $\chi_1(3\omega,t_a)$ -see \cite{Epa12}. Once $\chi_{3}(\omega,t_a)$ is known, the key quantity is, according to Eq. (\ref{eq1}), $X_{3}(\omega,t_a)$ defined as: 
\begin{equation}
X_{3}(\omega,t_a) \equiv \frac{\chi_{3}(\omega,t_a)}{Z(t_a)} \approx N_{corr}(t_a) {\cal H}(\omega \tau_{\alpha}(t_a))
\label{eq2}
\end{equation}

\begin{figure}[t]
\includegraphics[width=9cm,height=6.5cm]{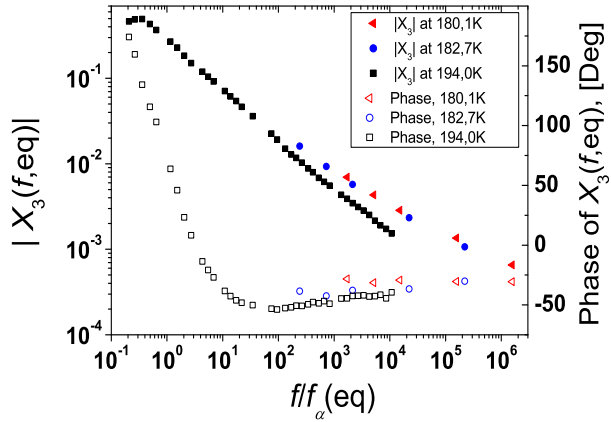} 
\caption{\label{fig1} (Color online) Equilibrium values of the modulus and the phase of $X_3$, measured in glycerol for three temperatures 
around $T_g \simeq 188\mathrm{K}$.}
\end{figure}

The equilibrium values of $X_{3,eq}(\omega,T)$, obtained after the end of aging, are plotted as a function of $f/f_{\alpha,eq}$ in Fig. \ref{fig1} for $T=180.1\mathrm{K}$ and for $T=182.7\mathrm{K}$ ($f_\alpha$ is defined as the peak frequency of $\chi''_{1,eq}(\omega,T)$, \cite{VFT}). For comparison, we also plot the equilibrium data at $194.1\mathrm{K} \approx T_g+6\mathrm{K}$ obtained in \cite{Cra10,Bru11} which shows that the qualitative trend already found above $T_g$ \cite{Cra10,Bru11} holds also below $T_g$, i.e. $\vert X_{3,eq} \vert$ for fixed $f/f_{\alpha,eq}(T)$ increases when $T$ decreases, \cite{Fbe}.

\begin{figure}[t]
\includegraphics[width=9cm,height=6.5cm]{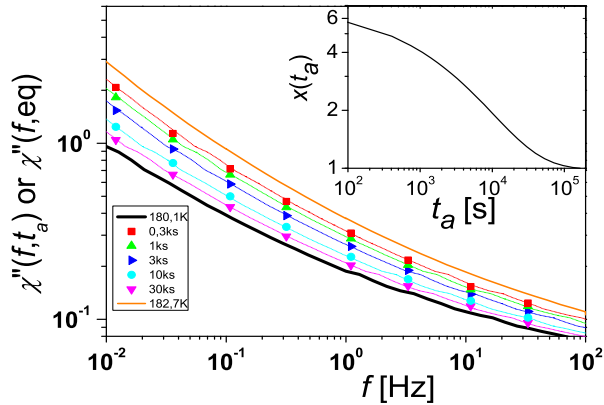} 
\caption{\label{fig2} (Color online) Aging of the out-of-phase susceptibility $\chi_1''(\omega,t_a)$ of glycerol at $T=180.1\mathrm{K}$ (filled symbols), in log-log coordinates. The thick (resp. thin) solid line correspond to the equilibrium spectrum at $T=180.1\mathrm{K}$ (resp. at $T=182.7\mathrm{K}$). 
The dotted lines superimposed to the filled symbols are obtained by translating horizontally the $180.1\mathrm{K}$ equilibrium spectrum by a factor $x(t_a)$ (see text). \textit{Inset:} Resulting curve for the scaling factor $x(t_a)=f_{\alpha}(t_a)/f_{\alpha,eq}$.} 
\end{figure}

\begin{figure}[t]
\includegraphics[width=9cm,height=6.5cm]{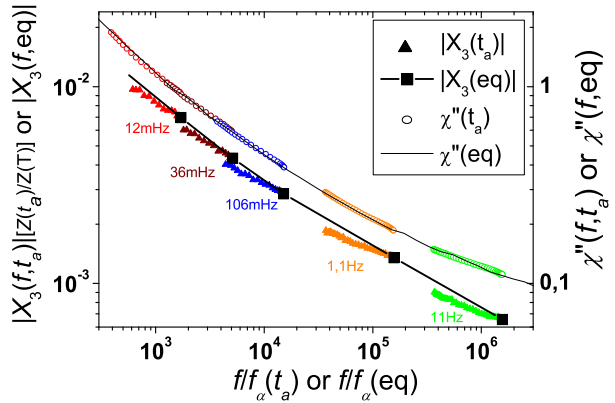} 
\caption{\label{fig3} (Color online) $T=180.1\mathrm{K}$. Solid curves: equilibrium quantities \textit{vs} $f/f_{\alpha,eq}$. Small symbols: aging 
quantities \textit{vs} $f/f_{\alpha}(t_a)$ -the value of $f$ labels each data set-. Note the collapse of the aging and equilibrium curves for $\chi_1''$ (right axis). This contrasts with 
what is observed for $\vert X_3 \vert$ (left axis) where the aging values are 
systematically \textit{below} the equilibrium ones. This reflects the \textit{increase} of $N_{corr}(t_a)$ during aging; see text, 
Eqs. (\ref{eq2},\ref{eq4}) and Fig. \ref{fig4}.} 
\end{figure}

\begin{figure}[t]
\includegraphics[width=9cm,height=6.5cm]{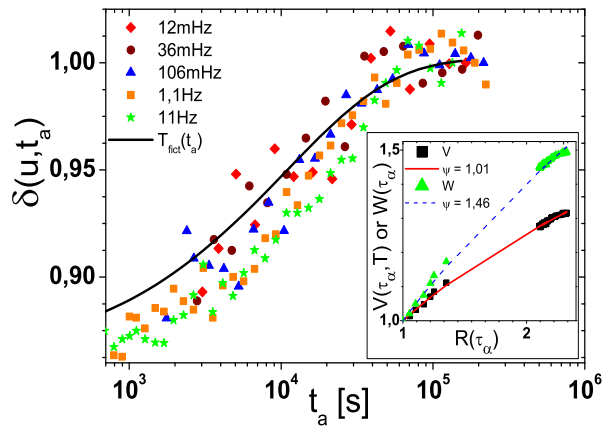} 
\caption{\label{fig4} (Color online) $T=180.1\mathrm{K}$. Values of $\delta(u,t_a)= Z(t_a) N_{corr}(t_a)/Z(T)N_{corr}(eq) \approx N_{corr}(t_a)/N_{corr}(eq)$ extracted from Fig. \ref{fig3}. $\delta(u,t_a)$ is found to be independent of frequency, as predicted by Eqs. (\ref{eq2},\ref{eq4}). The solid line is the estimation based on phenomenological fictive temperature $T_{fict}(t_a)$, see text. \textit{Inset:}  Estimation of the exponent $\psi$ (see text) using all aging and equilibrium data, starting from $204\mathrm{K}$ where $\tau_{\alpha,eq}=20\mathrm{ms}$. The $x$-axis is $R(\tau_{\alpha})=N_{corr}(\tau_{\alpha})/N_{corr,eq}(20\mathrm{ms})$ with $\tau_{\alpha}=\tau_{\alpha}(t_a)$ or $\tau_{\alpha,eq}$. For the $y$-axis, we chose  either $\Upsilon=\Upsilon_0$ which amounts to $V=v(\tau_{\alpha},T)/v(20\mathrm{ms},204\mathrm{K})$ and yields $\psi \approx 1$; or $\Upsilon=\kappa k_B T$ which amounts to $W=w(\tau_{\alpha})/w(20\mathrm{ms})$ and yields $\psi \approx 3/2$.} 
\end{figure}

{\it Scaling analysis.} During aging both $\vert \chi_1 \vert$ and $\vert X_3 \vert$ decrease for a given $\omega$, because the dielectric spectrum shifts
to lower frequencies \cite{Leh98,Yar03,Lun05,Shi05,Hec10,Ric10}. This is illustrated by the series of symbols on Fig. \ref{fig2}. 
For not too deep quenches \cite{Leh98}, such as ours, and for our limited range of $3.5$ decades in frequencies, the shape of the spectrum of $\chi_1''$ is not expected to change during aging except for an overall scaling factor. This is confirmed by  Fig. \ref{fig2} where the $\chi_1''(\omega,t_a)$ data for all frequencies and all ages can be very accurately reproduced by the equilibrium susceptibility 
$\chi_{1,eq}''(\omega,T=180.1\mathrm{K})$, up to a rescaling of the frequency by a factor $x(t_a)=f_{\alpha}(t_a)/f_{\alpha,eq}$ (see the dotted lines in Fig. \ref{fig2} obtained by adjusting the factor $x(t_a)$ for each $t_a$). We have also checked that our values of $x(t_a)$ are close to what is predicted by the ansatz introduced in Ref. \cite{Lun05}, \cite{Ans}.
We have checked that the very same $x(t_a)$ factor also allows us to rescale the $\chi_1'(\omega,t_a)$ data onto the equilibrium curve. 
Note that since the $\chi_1''(\omega,t_a)$ are not pure power-laws in frequency, horizontal and vertical shifts (in log-log) are not 
equivalent. Hence, the accurate rescaling of Fig. \ref{fig2} implies that the {\it amplitude} of $\chi_1''(\omega,t_a)$ does not depend on the age. A finer look at the  rescaling suggests that this amplitude is constant within a $1 \%$ uncertainty range, and if anything, {\it decreases} with age. This is important for the discussion of the $Z(t_a)$ factor in Eq. (\ref{eq1}) that includes $\Delta \chi_1(t_a)$ to which $\chi_1''(\omega,t_a)$ is proportional. 
To estimate the difference between $Z(t_a)$ and $Z(T)$ we invoke the ``fictive'' temperature $T_{fict}(t_a)$ \cite{NMT,Leh98,Lub04,Mos04} (not to be confused with the effective  temperature $T_{eff}$) defined such that $\chi_1''(\omega,t_a) = \chi_{1,eq}''(\omega, T_{fict}(t_a))$. This phenomenological recipe leads to $T_{fict}(t_a > 0.3\mathrm{ks}) < T+2\mathrm{K}$. By extrapolating the
 $T$ dependence of $\Delta \chi_1$ above $T_g$, we estimate that  $\Delta \chi_1(t_a)$ may {\it increase} by at most $1.1\%$ during aging. 
The order of magnitude is similar to the one suggested by the rescaling analysis above, albeit with an opposite sign. Similarly, we estimate that $a^3$ might \textit{decrease} by 
$\sim 0.2\%$ during aging. Finally, close to $180\mathrm{K}$, $T/T_{eff}(t_a)$ was found to increase in glycerol during aging, by $\sim 2\%$ according to \cite{Gri99}, but by at most $0.7\%$ according to the recent work of Ref. \cite{Sch11}. 
Altogether, we conclude that $Z(t_a)/Z(T)$ remains very close to unity, with a probably much overestimated maximum increase of $4 \%$ during aging. This effect is therefore smaller than the $\sim 12 \%$ increase of $N_{corr}(t_a)$ that we infer from our analysis below. 

Our central experimental result is summarized in Fig. \ref{fig3} where we now show both $\chi_1''(\omega,t_a)$ and $\vert X_{3}(\omega, t_a) \vert$ as a function of $f/f_{\alpha}(t_a)$ for $1\mathrm{ks} \leq t_a \leq 200\mathrm{ks}$. As expected from the results of Fig. \ref{fig2}, $\chi_1''(\omega,t_a)$ collapses very well onto the equilibrium curve. However, this collapse is {\it not} observed for $\vert X_{3}(\omega,t_a) \vert$ (Fig. \ref{fig3}, filled triangles and left axis). The rightmost points of these series of triangle correspond to the equilibrium values $\vert X_{3,eq}(\omega,T) \vert$ and are singled out as large black squares. The thick line joining these black squares is an interpolation that corresponds to the equilibrium value of $\vert X_{3,eq}(\omega,T) \vert$ for intermediate frequencies. At variance with the good superposition obtained 
for $\chi_1''$, Fig. \ref{fig3} reveals that, for a given $f/f_{\alpha}(t_a)$, the value of $\vert X_{3}(\omega,t_a) \vert$ is systematically \textit{below} the corresponding value at equilibrium. This is exactly what is expected from Eq. (\ref{eq2}): the ratio between these two values should 
be equal to $N_{corr}(t_a)/N_{corr,eq}(T)$, and should thus increase with age, precisely as observed in Fig. \ref{fig3}. Defining the vertical logarithmic distance $\delta$, in Fig. \ref{fig3}, as:
\begin{equation}
\delta (u, t_a) \equiv \frac{Z(t_a)X_{3}(u f_{\alpha}(t_a), t_a)}{Z(T)X_{3,eq}(u f_{\alpha,eq})} = \frac{Z(t_a)N_{corr}(t_a)}{Z(T)N_{corr,eq}(T)} ,
\label{eq4}
\end{equation}
we obtain the last equality if  Eq. (\ref{eq2}) holds, in which case $\delta (u, t_a)$ should be {\it independent} of frequency.
With $Z(t_a)/Z(T) \approx 1$ justified above,  we conclude that $\delta (u, t_a)$ 
directly measures $N_{corr}(t_a)/N_{corr,eq}(T)$. 

The values of $\delta(u, t_a)$ are plotted in Fig. \ref{fig4}, \cite{182K}. We indeed observe that, up to the precision of our measurements, $\delta(u,t_a)$
does not depend on  frequency. This is an important consistency test of our scaling assumption, Eqs. (\ref{eq1},\ref{eq2}). From Fig. \ref{fig4}, we deduce that $\delta(u,t_a)$ increases by $\approx 12\%$ when $t_a$ increases from $1\mathrm{ks}$ to $100\mathrm{ks}$. Since the ratio $Z(t_a)/Z(T)$ increases 
by at most $4\%$, we interpret the data of Fig. \ref{fig4} as giving the first experimental evidence that the size of the dynamically correlated 
clusters increases with the age in a glass former, see \cite{quench}. The increase of 
$N_{corr}(t_a)$ during aging can be approximately accounted for by extending the observation made in \cite{Cra10,Bru11}: the temperature dependence of $N_{corr,eq}$ deduced from non-linear susceptibility measurements can alternatively 
be obtained as: $ \partial N_{corr,eq} /\partial T \approx 1.5 \partial(T\chi_T) /\partial T$, where  $T\chi_T = T  \times max_{\omega} \vert \left[ \partial (\chi'_{1,eq}(\omega,T)/\Delta \chi_1) / \partial T \right] \vert $, see \cite{Bru11,Ber05,Ber07,note1}. We now surmise that this can be extended to the out-of-equilibrium regime by simply translating the $f_{\alpha}(t_a)$ dependence of Fig. \ref{fig2} in terms
of $T_{fict}(t_a)$ (see \cite{Lefevre}). 
This heuristic procedure leads to the solid curve shown in Fig. \ref{fig4}, which is indeed close to the values of $\delta(u,t_a)$ directly drawn from our experiments. This suggests that it might be possible to extend the theoretical work of \cite{Ber05,Ber07} to aging, and get  a simplified way of estimating $N_{corr}(t_a)$ using linear susceptibilities.

{\it Time and length scales.} Finally, we take advantage of the wide range of time scales over which the evolution of $N_{corr}$ has been measured to revisit one of the 
most crucial aspect of glassy dynamics, namely the relation between time and length scales. Within the Random First Order Transition theory 
 \cite{Lub06,Bir12}, one expects $\ln (\tau_{\alpha}/\tau_0) = \Upsilon \ell_{PS}^{\psi}/k_B T$, where $\tau_0$ is a microscopic time scale, $\Upsilon$ a typical molecular energy barrier, $\ell_{PS}$ the point-to-set correlation length \cite{Leiden,Bir12}, which sets the size of the clusters that must rearrange cooperatively for the system to relax, and $\psi$ the so-called barrier exponent. In Wolynes' version of RFOT, $\Upsilon = \kappa k_B T$ where $\kappa$ is a number that depends weakly on molecular details, and $\psi=3/2$ \cite{Lub06}. In order to compare with our results, one should postulate that the size of dynamically correlated clusters $N_{corr}$ is proportionnal to $\ell_{PS}^3$. This relation is not unreasonable, but sharp theoretical arguments are still lacking to relate unambiguously ``cooperative'' regions to ``dynamically correlated regions''. In any case, we collect all our past and present data in the inset of Fig. \ref{fig4}, where we plot  $v(\tau_{\alpha},T)=T\ln (\tau_{\alpha}/\tau_0)$ and $w(\tau_{\alpha})=\ln (\tau_{\alpha}/\tau_0)$ as a function of $N_{corr}$,
where $\tau_{\alpha}=\tau_{\alpha}(t_a)$ or $\tau_{\alpha,eq}$. In Fig. \ref{fig4}, $V$ and $W$ correspond respectively to $\Upsilon = \Upsilon_0$ independent of temperature and to $\Upsilon = \kappa k_B T$. We fix the value of $\tau_0$ to $10^{-14}\mathrm{s}$, leaving $\psi$ as the only free parameter. To our surprise, we find that the best choice for $\psi$ within the first hypothesis ($\Upsilon = \Upsilon_0$) is $\psi \approx 1$, which is the value found numerically in \cite{Cam10}, whereas in the second hypothesis ($\Upsilon = \kappa k_B T$), we find $\psi \approx 3/2$, as predicted by Wolynes \textit{et al.}! Our data is compatible with both hypotheses \cite{Par}, although slightly favoring the first one, in particular in the aging regime (see Fig. \ref{fig4}). Note that a factor $10$ on $\tau_0$  changes the value of $\psi$ by $\sim 10 \%$.

{\it Conclusion.} We have reported the first direct observation of the increase of $N_{corr}$ in the aging regime of a structural glass. For glycerol 
at $T=0.96 T_g$, our quench protocol yields an increase of $N_{corr}$ by $\sim 10\%$, which lasts $\sim 100\mathrm{ks}$ \cite{quench}. These results deepen our microscopic understanding of aging and give precious information about the relation between time and length scales in glasses. Our study opens a new path for studying aging in many other systems. It could also be extended to the more complicated thermal histories designed to probe the memory and rejuvenation effects \cite{Vin07,Yar03}. Monitoring the behaviour of $N_{corr}$ in these experiments should shed a new light on these phenomena.

We thank C. Alba-Simionesco, S. Nakamae, G. Tarjus, R. Tourbot for help and discussions. GB acknowledges financial support from the ERC grant NPRG-GLASS.

\begin{center}
{\bf SUPPLEMENTARY INFORMATION:}
\end{center}

For the sake of completness, we give here the main ingredients of the $3\omega$ measurements in the aging regime. \textit{Note that the quantity defined as  $\chi_3^{(3)}$ in Eq. (6) below has been noted $\chi_3$ in the main letter, to simplify the notations}. 

As explained in Refs. \cite{Thi08, Bru11}, when a field $E(t)$ is applied onto a dielectric liquid, the macroscopic polarisation $P$ can be expressed as : 

\begin{eqnarray}
\frac {P(t)}{\epsilon_0} &=& \int_{-\infty}^{\infty}{\chi_1(t-t')E(t')dt'} \\ \nonumber
\ &\ & + \iiint_{-\infty}^{\infty} {\chi_3(t-t'_1,t-t'_2,t-t'_3) E(t'_1) \times} \\ \nonumber 
\ &\ & \times E(t'_2)E(t'_3)dt'_1dt'_2dt'_3 + ..., 
\label{eqS1}
\end{eqnarray}

where the function $\chi_1(t)$ corresponds to the experimental macroscopic linear response 
while $\chi_3(t_1,t_2,t_3)$ is the experimental macroscopic nonlinear response. This expression is valid as long as the non linear terms are small, i.e. as long as, for any integer $k$, $\vert \chi_{2k+1}E^{2k+1} \vert \ll \vert \chi_{2k-1}E^{2k-1} \vert$. This is why Eq. (5) is restricted to the cubic response, and neglects higher order terms.

It is shown in ref. \cite{Thi08}, that  for a field $E(t)= E \cos(\omega t)$ one gets, from Eq. (5): 
\begin{eqnarray}               
\frac{P(t)}{\epsilon_0} &=& E\left|\chi_1\right| \cos(\omega t - \delta_1) \\ \nonumber 
\                       &\ & +3/4 E^3\left|\chi_{3}^{(1)}\right| \cos(\omega t - \delta_{3}^{(1)}) \\ \nonumber
\                       &\ & + 1/4E^3\left|\chi_3^{(3)}\right| \cos(3\omega t - \delta_3^{(3)})+...  
\label{eqS2} 
\end{eqnarray}

The time dependent polarisation amounts to an electrical current given by 
\begin{equation}
I(t) = S \frac{\partial P(t)}{\partial t}
\label{eqS3}
\end{equation}
where $S$ is the surface of the electrodes. Inserting Eq. (6) into Eq. (7), one finds that $I(t)$ is the sum of a current 
$I(1\omega,t)$, oscillating at the fundamental frequency, and of a current $I(3\omega,t)$ oscillating at $3\omega$. As for the field range 
$E \le 3$MVrms/m involved in our experiments the condition mentionned above  $\vert \chi_{2k+1}E^{2k+1} \vert \ll \vert \chi_{2k-1}E^{2k-1} \vert$
 is satisfied, one gets firstly that $\vert I(3\omega,t) \vert \ll \vert I(1 \omega, t) \vert$; secondly that the value of $I(1\omega, t)$ is 
 fully dominated by $\chi_1$ and thus $I(1 \omega, t)$ can be, to a very good approximation, analysed by using the usual framework of 
 complex admittance ${\cal Y}(\omega)$. The small third harmonics current is given by:

\begin{equation}
I(3\omega, t)= \frac{3}{4} \epsilon_0 S \omega \chi_3^{(3)}(\omega) E^3 \cos(3 \omega t +\frac{\pi}{2} - \delta_3^{(3)})
\label{eqS4}
\end{equation}

\begin{figure}[t]
\includegraphics*[width=6.0cm,height=9.0cm,angle=-90]{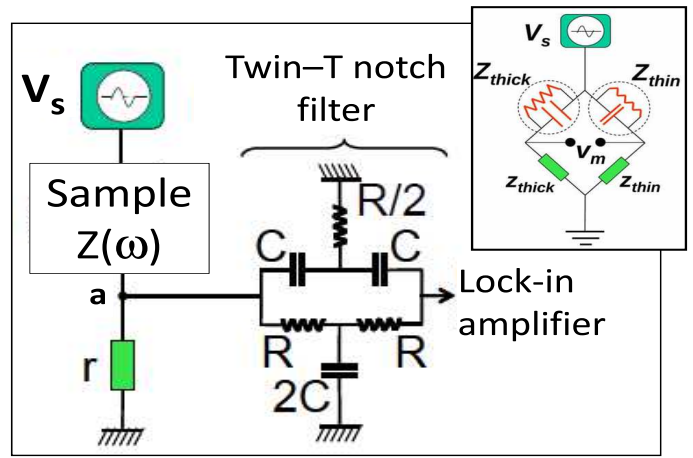}
\caption{\textit{Main figure}: Electronic setup used for the measurements of $\chi_{3}^{(3)}(\omega, t_a)$ in the aging regime. Note that the quantity defined as  $\chi_3^{(3)}$ in Eq. (6) of this Supplementary Material has been noted $\chi_3$ in the main letter, to simplify the notations. The ``twin-T'' notch-filter damps the response at $1 \omega$ by a factor larger than $10^4$. \textit{Inset}: Two samples bridge, see ref. \cite{Thi08}.}
\label{figS1}
\end{figure}

This quantity is so small that carefully designed electronic setups must be used to avoid to mix the sought $I(3\omega,t)$ with the 
nonlinear imperfections of the voltage source and of the voltage amplifier \cite{Thi08}. It was shown in Ref \cite{Thi08} that two kinds of 
setups can be used: either a ``two samples bridge'' involving two samples of different thicknesses -see the inset of  Fig. \ref{figS1}-; or a ``twin-T notch filter'', see the main part of Fig. \ref{figS1}. 

The two samples bridge is a technique measuring a differential voltage 
$V_m  \equiv V_{thin}-V_{thick}$ and 
relying on the ``balancing relation'' ensuring that $V_m(1 \omega) =0$: this happens provided one has $z_{thick} {\cal Y}_{thick} = z_{thin} {\cal Y}_{thin}$ where ${\cal Y}$ is the admittance of one sample and $z$ 
is the impedance relating the sample  to the ground. An important feature of the two samples bridge is that once the balancing condition is met 
at $1 \omega$ it is also met at any other frequency. Thus the balancing condition enables, at the same time, to suppress the contribution coming
 from the nonlinear character of the input voltage amplifier -since $V_{thin}(1 \omega)-V_{thick}(1 \omega) = 0$-; and to suppress 
 the $3 \omega$ spurious 
 component of the source -since 
 the linear response of the samples cancels at any frequency-. The two samples bridge is thus the most efficient technique to get $I(3\omega,t)$,
  \textit{provided one has a way to check} that the condition $V_{thin}(1 \omega)-V_{thick}(1 \omega) = 0$ remains true during the 
  $I(3\omega,t)$ acquisition. 
  This condition might indeed not remain true in the case where some uncontrolled slight disymetry between the two samples happens, such as the 
  one resulting from a slight difference in the temperature of the two samples. Fortunately, 
  when measuring at equilibrium, one varies the field $E$: one thus checks all along the $3\omega$ acquisitions that $I(3\omega,t)$ is accurately 
  -up to $\pm 1\%$- proportional to $E^3$, which ensures that the condition $V_{thin}(1 \omega)-V_{thick}(1 \omega) = 0$ is met -enough- during all 
  the acquisitions. However, in the aging case, one monitors $I(3\omega,t)$ as a function of the age $t_a$, for a constant $E$. Thus, 
  if some violation 
  of the condition $V_{thin}(1 \omega)-V_{thick}(1 \omega) = 0$ happened, it might pollute the age dependence of the $3 \omega$ response of the samples, and
   we would have no way to correct this imperfection. 
   
   This is why we have decided to work with only one sample and to use 
   the ``Twin-T notch filter'' -see Fig. \ref{figS1}-: its transmission coefficient 
  ${\cal T}_{filt}(\omega)$ is smaller than $10^{-4}$ for the frequency $f_0 = 1/(2 \pi R C)$ and of order $1$ for $3 f_0$. Therefore, by choosing the components $R, C$ so as to set $2 \pi f_0 = 1 \omega$, we are \textit{absolutely sure} that the $1 \omega$ voltage at the input of the Lock-in amplifier is small enough during the $3 \omega$ acquisitions in the aging regime. Besides, by setting the voltage source to one of the few values where the DS360 voltage source is nearly perfectly harmonic, one gets a setup where the $3\omega$ spurious contribution of the source is nearly negligible -this very little spurious contribution can, of course, be easily measured and subtracted from the measured signal-. This is why we have choosen the twin T notch filter for our $3\omega$ measurements in the aging regime.

At $3 \omega$ the sample is equivalent to a pure current source $I(3 \omega, t)$ with an impedance ${\cal Z}(3\omega) = 1/[{\cal Y}(3\omega)]$ placed in parallel. Neglecting for simplicity any remaining spurious contribution of the DS360 source, one gets for the voltage $V_{meas}(3\omega,t)$ measured by the lock-in amplifier: 

\begin{equation}
V_{meas}(3\omega, t) = \alpha (3 \omega) {\cal Z}(3 \omega) I(3 \omega,t)
\label{eqS5}
\end{equation}

where $\alpha(3\omega)$ is the global transmission coefficient between the sample and the Lock-in. In the simplest case 
where $(r, \vert {\cal Z}\vert)  \ll (R, 1/[C 3 \omega])$ the coefficients multiply and one 
gets $\alpha(3 \omega) = {\cal T}_{filt}(3\omega) \times r/[r+{\cal Z}(3\omega)]$. 
In any case, $\alpha(3 \omega)$ is directly measured by setting the fundamental angular frequency of the source to $\Omega \equiv 3 \omega$, and
 by using $\alpha(\Omega) \equiv  V_{meas}(\Omega)/ V_{source}(\Omega)$. 

Eq. (9) is written at equilibrium, when all the involved quantities no longer depend on the age $t_a$. At equilibrium, we have of 
course carefully checked, both at $T=180.1$K and at $T=182.7$K, that $V_{meas}(3\omega,t)$ is proportionnal to the cube of the voltage source 
$V_s^3$. In the aging case, all the susceptibilities of the sample -linear and nonlinear- depend on the age $t_a$. 
As a result, all the quantities involved in Eq. (9) depend on the age $t_a$. This is why, by repeating for 
each quantity the very same $T$ quench, we have separately measured the age dependence of $I(3\omega,t,t_a)$ but also of ${\cal Z}(3 \omega,t_a)$, 
${\cal Z}(1 \omega,t_a)$, $\alpha (3 \omega,t_a)$, and of $V_{appl} (1 \omega,t,t_a) = V_s(1\omega,t) - V_a(1\omega,t,t_a)$ 
where $V_{appl}(1\omega,t,t_a)$ is the voltage applied onto the sample at age $t_a$ and at time $t$ -see Fig. \ref{figS1}-. 
We have made all the possible consistency 
checks -for example, given $R,C, {\cal Z}(1 \omega, t, t_a)$ one can \textit{predict} the age dependence of $V_{appl} (1 \omega,t,t_a)$-. 
A single $T$ quench lasts $200$ks at $180.1$K, and for each of the $5$ different frequencies ranging from $12$mHz to $11$Hz, we have measured the 
age dependence of the $5$ different quantities mentionned above. As a result, with all the cross-checks, the data acquisition, for 
 $180.1$K and $182.7$K altogether, took a bit more than $10^4$ks, i.e. a bit more than 4 months.


\begin{thebibliography}{99}
\bibitem{Tri} Following Ref. \cite{Bru11}, as $f \ge 10f_{\alpha}$, the trivial contribution to Eq. (\ref{eq1}), not related to $N_{corr}$, is  negligible here.
\bibitem{VFT} $f_{\alpha,eq}(T<T_g)$ was estimated by extrapolating the VFT law obtained for $f_{\alpha,eq}(T>T_g)$: as some deviation to VFT behavior might arise below $T_g$ \cite{Shi05,Lub04}, this introduces some uncertainty. Thus Fig. \ref{fig1} is not a strict comparison between the data above and below $T_g$.
\bibitem{Fbe} The frequencies reported in Fig. \ref{fig1} belong to the $\alpha$ relaxation peak for which Eq. (\ref{eq1}) was originally derived. In glycerol the small excess wing indeed corresponds to a typical frequency  larger than $10^5 f_{\alpha,eq}$ close to $T_g$ \cite{Lun02}.
\bibitem{Ans} Note however that this ansatz requires the knowledge of $f_{\alpha}(t_a=0)$ and $f_{\alpha,eq}$, which introduces some uncertainty, see \cite{VFT}.
\bibitem{182K} At $T=182.7\mathrm{K}$, as expected, we find weaker aging effects: aging lasts $\simeq 20\mathrm{ks}$, and yields a relative increase of $N_{corr}$ with $t_a$ typically $2-3$ times smaller than that of Fig. \ref{fig4}. These results are thus consistent with those at $180.1\mathrm{K}$.
\bibitem{quench} Extrapolating the $T$ dependence of $N_{corr,eq}(T)$ measured above $T_g$ \cite{Cra10,Bru11}, one estimates that the quench from $T_g +8\mathrm{K}$ to $T_g-8\mathrm{K}$ corresponds to a doubling of $N_{corr,eq}$. The $\sim 10\%$ increase of $N_{corr}(t_a)$ reported here is thus the long-time ``tail'' part of this increase (the first $90\%$ increase cannot be measured because it takes place during the quench).
\bibitem{note1} Here we use the $T$-dependence of $N_{corr}$ obtained in Ref. \cite{Bru11} where the role of the ``trivial'' nonlinear response 
has been thoroughly eliminated. This had not been done in Ref. \cite{Cra10}, which explains why the $T$-dependence of $N_{corr}$ was underestimated in Ref. \cite{Cra10}.
\bibitem{Par} The fit  $\tau_{\alpha} \propto N_{corr}^z$ proposed in \cite{Parisi,Castillo,nandi} is of similar quality but leads, here, to unreasonably large values of $z \sim 20$.
\end{thebibliography}
\end{document}